\documentclass[pre,twocolumn]{revtex4}
\usepackage{epsfig}
\usepackage{bm}

\begin{document}

\title{Helicity, low-dimensional distributed chaos and scaling in MHD turbulence}

\author{A. Bershadskii}

\affiliation{
ICAR, P.O. Box 31155, Jerusalem 91000, Israel
}

\begin{abstract}

  It is shown that the second-order moment of helicity distribution (the Levich-Tsinober invariant of inviscid hydrodynamics) can be preserved as an adiabatic invariant in magnetohydrodynamics. The distributed chaos approach and inertial range phenomenology have been developed for the magnetohydrodynamic processes dominated by this adiabatic invariant, and some consequences of the theoretical consideration have been compared with results of direct numerical simulations and measurements in the solar wind. Helical scaling in interstellar MHD turbulence has been also briefly discussed in this context.

\end{abstract}

\maketitle

\section{Introduction}

  The multiplicity of regimes in fluids dynamics with active scalars and vectors (e.g. in the thermal convection and magnetohydrodynamics) is related to multiplicity of the invariants in such systems. 

  For instance, the helicity
$$
H = \int  h({\bf x},t)~ d{\bf x},  \eqno{(1)}
$$
where the helicity density 
$$
h ({\bf x},t) =   {\bf u} ({\bf x},t) \cdot  {\boldsymbol \omega} ({\bf x}, t),   \eqno{(2)}
$$   
${\bf u} ({\bf x},t)$ is the velocity field and ${\boldsymbol \omega} ({\bf x}, t)= [\nabla \times {\bf u}]$ is the vorticity field, is a fundamental inviscid hydrodynamic invariant (see for a review Ref. \cite{mt}). In magnetohydrodynamics (MHD), however, the helicity Eq. (1) is not invariant and the magnetic helicity and the cross-helicity
$$
H_{cr} = \int {\bf u}({\bf x},t)\cdot {\bf B}({\bf x},t) d{\bf x},  \eqno{(3)}
$$
where ${\bf B}({\bf x},t)$ is the magnetic filed, take its place (see the Ref. \cite{mt} and references therein).\\

  It was shown (see Refs. \cite{mt},\cite{lt} and references therein) that under certain rather general conditions the second order moment of helicity density is also an inviscid hydrodynamic invariant (the Levich-Tsinober invariant), and it will be shown in present paper that this invariant (unlike the helicity itself) can be preserved also in the magnetohydrodynamics as an adiabatic invariant. \\
  
 The MHD regimes dominated by this adiabatic invariant have been studied in present paper using the distributed chaos approach and inertial range phenomenology, and compared with the results of direct numerical simulations (DNS) and measurements in the solar wind. \\

\section{Second order moments of helicity distribution}  

The magnetohydrodynamics of incompressible fluids is described by equations:
$$
 \frac{\partial {\bf u}}{\partial t} = - {\bf u} \cdot \nabla {\bf u} 
    -\frac{1}{\rho} \nabla {\cal P} - {\bf F}_L + \nu \nabla^2  {\bf u} + {\bf f_u} \eqno{(4)}
$$
$$
\frac{\partial {\bf b}}{\partial t} = \nabla \times ( {\bf u} \times
    {\bf b}) +\eta \nabla^2 {\bf b} +  {\bf f_b} \eqno{(5)} 
$$
$$ 
\nabla \cdot {\bf u}=0, ~~~~~~~~~~~\nabla \cdot {\bf b}=0,  \eqno{(6)}
$$
where the normalized magnetic field ${\bf b} = {\bf B}/\sqrt{\mu_0\rho}$ has the same dimension as velocity (the Alfv\'enic units), ${\bf F}_L = [{\bf b} \times (\nabla \times {\bf b})]$ is the  Lorentz force, ${\bf f}_u$ and ${\bf f}_b$ are the external forcing functions (in the DNS these functions are usually concentrated in the large scales only). \\
 
 Dynamic equation for mean helicity can be readily obtained from the Eq. (4) and in the inviscid case  ($\nu =0$) has the form
$$
\frac{d\langle h \rangle}{dt}  = 2\langle {\boldsymbol \omega} \cdot (- {\bf F}_L + {\bf f_u})  \rangle \eqno{(7)} 
$$ 
where $\langle ... \rangle$ is an average over volume.
 
  One can see from the Eq. (7) that the helicity is not an inviscid MHD invariant. However, if the correlation $\langle {\boldsymbol \omega} \cdot (- {\bf F}_L + {\bf f_u}) $ is considerable at large scales and it quickly becomes negligible for the smaller scales in a turbulent environment, then the second order moment of the helicity distribution (the Levich-Tsinober invariant of the inviscid hydrodynamics \cite{lt}) can be an inviscid quasi-invariant of the magnetohydrodynamics as well. To show this the volume of motion should be divided into the cells $V_j$ (with boundary conditions ${\boldsymbol \omega} \cdot {\bf n}=0$ on the surfaces of the cells - $S_j$), which move with the fluid \cite{mt}. For the cells with the small enough spatial scales (so that the correlation $\langle {\boldsymbol \omega} \cdot (- {\bf F}_L + {\bf f_u})  \rangle$ over the cell is negligible) the helicity density, averaged over the cell, can be considered as an adiabatic invariant in an inertial range of scales. We can use the following definition of the second order moment \cite{mt}
$$
I = \lim_{V \rightarrow  \infty} \frac{1}{V} \sum_j H_{j}^2  \eqno{(8)}
$$
with 
$$
H_j = \int_{V_j} h({\bf x},t) ~ d{\bf x} \eqno{(9)}
$$
%%%%%%%%%%%%%%% 1 %%%%%%%%%%%%%%%%%%
\begin{figure} \vspace{-1.6cm}\hspace{-1.5cm}\centering
\epsfig{width=.53\textwidth,file=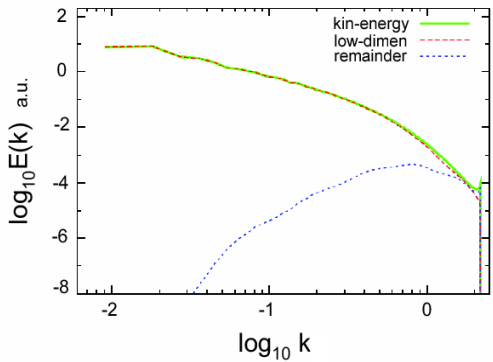} \vspace{-6.7cm}
\caption{ Kinetic energy spectra with the split on coherent (low-dimensional) and incoherent (remainder) parts. } 
\end{figure}
%%%%%%%%%%%%%%%%%%%%%%%%%%%%%%%%%%%  
%%%%%%%%%%%%%%% 2 %%%%%%%%%%%%%%%%%%
\begin{figure} \vspace{-1.2cm}\hspace{-1.5cm}\centering
\epsfig{width=.52\textwidth,file=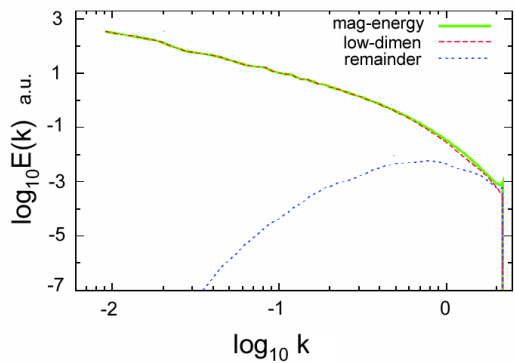} \vspace{-5.9cm}
\caption{ Magnetic energy spectra with the split on coherent (low-dimensional) and incoherent (remainder) parts. } 
\end{figure}
%%%%%%%%%%%%%%%%%%%%%%%%%%%%%%%%%%%
 For a strong MHD turbulence we can expect that the cells with the quasi-invariant $H_j^2$ determine the sum in the Eq. (8) (cf. Ref. \cite{bt} and references therein) and, as a consequence, $I$ is a quasi-invariant of the inviscid MHD dynamics and an adiabatic invariant in an inertial range of the viscid MHD turbulence.\\

   It should be also noted that in the non-dissipative case dynamics of the magnetic field ${\bf b}$ is described by equation
$$
\frac{\partial {\bf b}}{\partial t} = \nabla \times ( {\bf v} \times
    {\bf b})   \eqno{(10)}
$$    
that is similar to the equation describing inviscid dynamics of vorticity ${\boldsymbol \omega}$ \cite{mt}.  Therefore, the analogous consideration  can be applied to the second order moment of the cross-helicity density $h_{cr} = {\bf v}({\bf x},t)\cdot {\bf b}({\bf x},t)$ (replacing ${\boldsymbol \omega}$ by ${\bf b}$)
$$
I_{cr} = \lim_{V \rightarrow  \infty} \frac{1}{V} \sum_j H_{cr,j}^2,  \eqno{(11)}
$$
where
$$
H_{cr,j} = \int_{V_j} {\bf v}({\bf x},t)\cdot {\bf b}({\bf x},t) d{\bf x}  \eqno{(12)}
$$
Hence, the second order moment of the cross-helicity density $I_{cr}$ is an adiabatic invariant in an inertial range of the dissipative MHD turbulence.\\

  Taking into account the adiabatic invariance of the second order moment of helicity density $I$ we can obtain a relationship between characteristic velocity $u_c$ and characteristic wavenumber $k_c$ in an inertial range using dimensional considerations  
$$
u_c \propto I^{1/4} k_c^{1/4}  \eqno{(13)}
$$

   Analogously for the cross-helicity dominated case:
$$
u_c \propto I_{cr}^{1/4} k_c^{3/4},  \eqno{(14)}
$$  

  Since in the Alfv\'enic units ${\bf b}$ has the same dimensionality as ${\bf u}$ the same dimensional considerations result in
$$
b_c \propto I^{1/4} k_c^{1/4}  \eqno{(15)}
$$ 
for the helicity dominated case and in
$$
b_c \propto I_{cr}^{1/4} k_c^{3/4},  \eqno{(16)}
$$ 
for the cross-helicity dominated case.
 
\section{Distributed chaos}

   In fluid dynamics and in plasmas the exponential spectra
$$
E(k) \propto \exp-(k/k_c)  \eqno{(17)}
$$
are often appear at the onset of turbulent dynamics dominated by the deterministic chaos (see, for instance, Refs. \cite{mm},\cite{kds} and references therein).  The further development of turbulent dynamics usually results in fluctuations of the parameter $k_c$ and one has to use an ensemble averaging 
$$
E(k) \propto \int_0^{\infty} P(k_c) \exp -(k/k_c)dk_c  \eqno{(18)}
$$    
in order to obtain the turbulent spectra. Therefore, an estimation of the probability distribution $P(k_c)$ is crucial for this purpose. \\

  A natural generalization of the exponential spectra Eq. (17) is the stretched exponential spectrum 
$$
E(k) \propto \exp-(k/k_{\beta})^{\beta}  \eqno{(19)}
$$
 Then, the problem is to find the parameter $\beta$ in the Eq. (19). For the stretched exponential spectra Eq. (19) an asymptotic of the $P(k_c)$ at large $k_c$ can be readily obtained form comparison of the Eqs. (18) and (19) \cite{jon}  
$$
P(k_c) \propto k_c^{-1 + \beta/[2(1-\beta)]}~\exp(-\gamma k_c^{\beta/(1-\beta)}) \eqno{(20)}
$$  
 If we will write the relationships Eqs. (13) and (14) in a general form  
$$
u_c \propto  k_c^{\alpha}   \eqno{(21)}
$$
and assume that $u_c$ has a Gaussian distribution with zero mean \cite{my} a relationship 
$$
\beta = \frac{2\alpha}{1+2\alpha}  \eqno{(22)}
$$
is following from the Eqs. (20) and (21). Then for the helicity dominated turbulence the Eq. (13) provides $\alpha = 1/4$ and from the Eq. (22) we obtain $\beta = 1/3$, i.e the kinetic energy spectrum
$$
E(k) \propto \exp-(k/k_{\beta})^{1/3}  \eqno{(23)}
$$
whereas for the cross-helicity dominated turbulence the Eq. (14) provides $\alpha = 3/4$ and from the Eq. (22) we obtain $\beta = 3/5$, i.e. the kinetic energy spectrum
$$
E(k) \propto \exp-(k/k_{\beta})^{3/5}  \eqno{(24)}
$$
Analogous magnetic energy spectra are following from the Eqs. (15) and (16) respectively. \\

  It should be noted that the second order moments of the helicity (cross-helicity) distribution $I$ ($I_{cr}$) can be substantial even in the case of negligible {\it mean} helicity (cross-helicty). Therefore, it is interesting to compare this approach with the approach of the Ref. \cite{b2} where the mean cross-helicty plays a crucial role.

\section{Direct numerical simulations of the MHD turbulence}

  In Ref. \cite{farge} results of direct numerical simulations of incompressible MHD turbulence with zero mean magnetic field in a periodic domain were reported. The random external solenoidal forces were concentrated in low wavenumbers: $k < 2.5$ (see the Refs. \cite{farge},\cite{ya} for more details). The forcing did not generate mean magnetic and cross-helicity (the initial magnetic and cross-helicity were negligible). The initial kinetic and magnetic energy spectra $E_0^u(k)=E_0^b(k) = C k^4 \exp-(k^2/2)$. The Taylor-Reynolds number based on velocity $R_{\lambda} = 159$.\\

   A threshold method of orthogonal wavelets was used by the authors of the Ref. \cite{farge} in order to split the current density and vorticity fields into incoherent (with many degrees of freedom) and coherent (with a few percent of the degrees of freedom only) parts. Figures 1 and 2 show (in the log-log scales) the kinetic and magnetic energy spectra obtained in this simulation. The spectral data were taken from Figs. 5a and 5b of the Ref. \cite{farge}. One can see that the energy spectra corresponding to the coherent (low-dimensional) part are slightly different from the entire spectra for the small scales (large $k$) only. \\
   
   Figures 3 and 4 show the kinetic and magnetic energy spectra with the dashed lines indicating the stretched exponential law Eq. (23). Comparison with the Figs. 1 and 2 allows us to conclude that the helical low-dimensional distributed chaos  dominates the both kinetic and magnetic spectra in this DNS (see Ref. \cite{l} about relation between the Levich-Tsinober invariant and coherent states in turbulence).\\

%%%%%%%%%%%%%%% 3 %%%%%%%%%%%%%%%%%%
\begin{figure} \vspace{-1.5cm}\centering
\epsfig{width=.45\textwidth,file=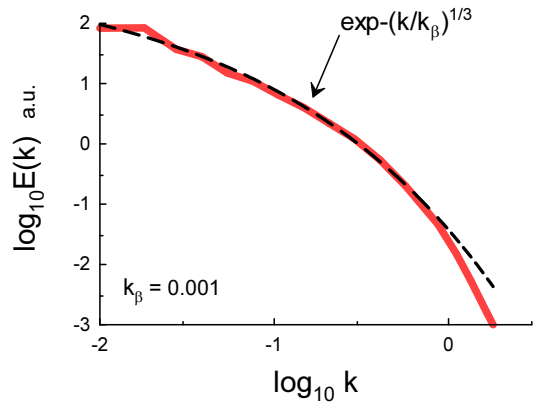} \vspace{-4.5cm}
\caption{The kinetic energy spectrum with the dashed line indicating the stretched exponential Eq. (23).} 
\end{figure}
%%%%%%%%%%%%%%%%%%%%%%%%%%%%%%%%%%%   
%%%%%%%%%%%%%%% 4 %%%%%%%%%%%%%%%%%%
\begin{figure} \vspace{-0.5cm}\centering
\epsfig{width=.45\textwidth,file=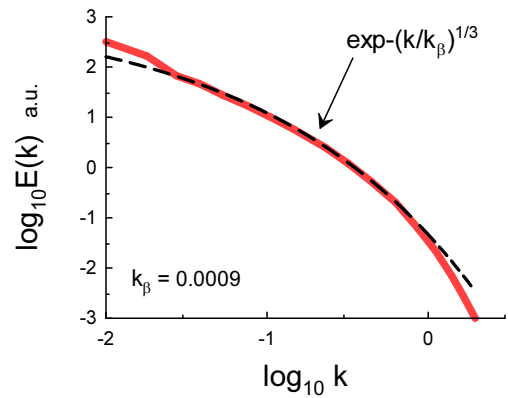} \vspace{-4.7cm}
\caption{The magnetic energy spectrum with the dashed line indicating the stretched exponential Eq. (23).} 
\end{figure}
%%%%%%%%%%%%%%%%%%%%%%%%%%%%%%%%%%%  

  In Ref. \cite{mpm} direct numerical simulations of decaying MHD turbulence with a superposition of deterministic (a Beltrami flow, cf. the Ref. \cite{l})  and random parts as initial conditions were performed. The deterministic part of the initial conditions consists of the ABC flows \cite{arn} in the large-scale wavenumber range $1 < k < 3$, whereas the random part is represented by small-scale Gaussian fluctuations. The initial mean cross-helicity is negligible.
  
  Figure 5 shows the magnetic energy spectrum observed at $t=1.6$ (the max. time of the system's evolution shown in the Fig. 2b of the Ref. \cite{mpm} where the spectral data were taken from). The dashed line indicates the stretched exponential law Eq. (23).\\
  
  It should be noted that the both Refs. \cite{farge} and \cite{mpm} emphasized crucial role of the current and vorticity sheets in the MHD turbulence.
  \\

   It is shown in recent Ref. \cite{bsb} that significant large-scale magnetic fields can be generated by helical turbulence in near incompressible conducting fluids despite the presence of strong small-scale dynamo effects. In the DNS reported in the Ref. \cite{bsb} a helical $\delta$-correlated in time forcing generates vortical motions in a large-scale wavenumber range around $k_f =4$. The Reynolds number $Re = 3300$ and the magnetic Prandtl number $Pr_m = 0.1$. \\
   
   Figures 6 and 7 show the kinetic and magnetic energy spectra with the dashed lines indicating the stretched exponential law Eq. (24). The spectral data were taken from Fig. 1 (final spectra) of the Ref. \cite{bsb}. The dotted vertical arrows indicate position of the scale $k_{\beta}$.

%%%%%%%%%%%%%%% 5 %%%%%%%%%%%%%%%%%%
\begin{figure} \vspace{-1.5cm}\centering
\epsfig{width=.45\textwidth,file=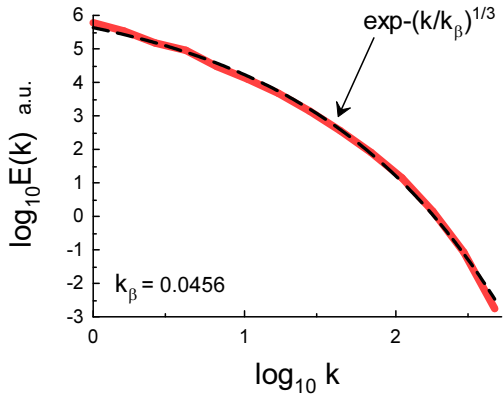} \vspace{-4.1cm}
\caption{Magnetic energy spectrum in the decaying MHD turbulence.} 
\end{figure}
%%%%%%%%%%%%%%%%%%%%%%%%%%%%%%%%%%%  
 %%%%%%%%%%%%%%% 6 %%%%%%%%%%%%%%%%%%
\begin{figure} \vspace{-1.3cm}\centering
\epsfig{width=.45\textwidth,file=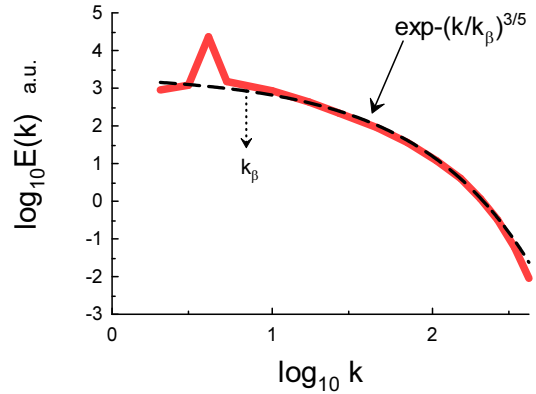} \vspace{-4.3cm}
\caption{Kinetic energy spectrum.} 
\end{figure}
%%%%%%%%%%%%%%%%%%%%%%%%%%%%%%%%%%% 
%%%%%%%%%%%%%%% 7 %%%%%%%%%%%%%%%%%%
\begin{figure} \vspace{-0.3cm}\centering
\epsfig{width=.45\textwidth,file=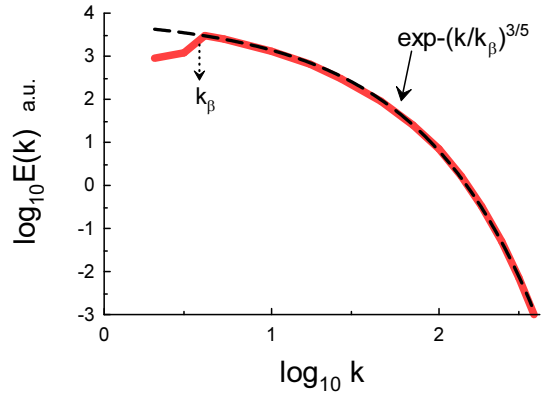} \vspace{-4.1cm}
\caption{Magnetic energy spectrum.} 
\end{figure}
%%%%%%%%%%%%%%%%%%%%%%%%%%%%%%%%%%% 

\section{Solar wind} 

\subsection{Helical distributed chaos}

  In the recent paper \cite{bpk} the authors use the Parker Solar Probe in situ measurements at radial distances between $35.7R_\odot$ and $41.7R_\odot$ from the center of the Sun to compute power spectra of the bulk velocity and magnetic fields in the solar wind (during the time period between November 3 and 9, 2018). They isolated the switchback intervals (corresponding to the folded magnetic field lines) from the nonswitchback intervals (those that are mainly following the Parker spiral field) of the measured signals. The power spectra were computed using the Fourier transform of  the corresponding conditioned correlation functions. The power spectra for the switchback  and nonswitchback intervals are shown in Figs. 8 and 9 respectively. The spectral data were taken from Fig. 4 of the Ref. \cite{bpk}. The noise level in the velocity power spectrum is shown using vertical lines.\\

%%%%%%%%%%%%%%% 8 %%%%%%%%%%%%%%%%%%
\begin{figure} \vspace{-0.7cm}\centering
\epsfig{width=.45\textwidth,file=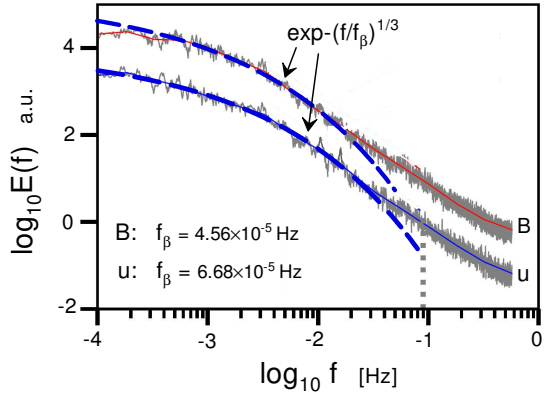} \vspace{-3.8cm}
\caption{Power spectrum of the bulk velocity and magnetic fields for the switchback intervals in the near-Sun solar wind.} 
\end{figure}
%%%%%%%%%%%%%%%%%%%%%%%%%%%%%%%%%%% 
%%%%%%%%%%%%%%% 9 %%%%%%%%%%%%%%%%%%
\begin{figure} \vspace{-0.3cm}\centering
\epsfig{width=.45\textwidth,file=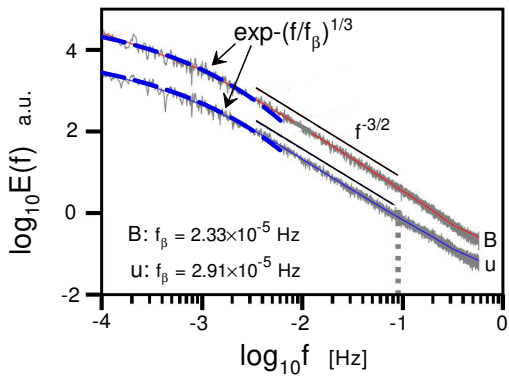} \vspace{-3.7cm}
\caption{As in the Fig. 8 but for the the nonswitchback intervals. }
\end{figure}
%%%%%%%%%%%%%%%%%%%%%%%%%%%%%%%%%%% 

  In the Figs. 8 and 9 the power spectra are plotted against the frequency (f) in the spacecraft frame. The mean solar wind velocity $|\langle {\bf V} \rangle|$ (in the spacecraft frame) is usually considerably larger than the corresponding velocity fluctuations. Therefore, the Taylor "frozen-in" hypothesis can be applied in this case (see, for instance, Refs. \cite{hb}), i.e the measured by a probe temporal dynamics reflects the spatial structures convected past the probe by the mean velocity. Hence the corresponding frequency spectra reflect the wavenumber ones with the replacement $k \simeq 2\pi f/ |\langle {\bf V} \rangle|$. The dashed curves in Figs. 8 and 9  indicate the stretched exponential law Eq. (23).

 \subsection{Helical scaling}

  In hydrodynamics an analogy of the Komogorov phenomenology 
$$
E(k) \propto \varepsilon^{2/3} k^{-5/3} \eqno{(25)}
$$
(where $\varepsilon =|d\langle {\bf u}^2 \rangle/dt|$ is energy dissipation rate) was applied to helicity in an inertial range of scales \cite{fr} (see also Ref. \cite{bt})
$$
E(k) \propto \varepsilon_h^{2/3} k^{-7/3}  \eqno{(26)}
$$
where $\varepsilon_h =|d\langle h \rangle/dt|$.\\
 
  Since in the magnetohydrodynamics the mean helicity is not an inviscid invariant one cannot expect that this phenomenology will work in the magnetohydrodynamics (though in a very strong external magnetic field it can be applied to the case of quasi two-dimensional MHD turbulence \cite{bkt}).  However, if the second order moment of helicity $I$ is still an adiabatic MHD invariant in the inertial range of scales  (see Section II) one can try to generalise the Kolmogov phenomelogy on this case as well. Unlike energy or helicity the $I$-invariant is not a quadratic invariant. Therefore, the $\varepsilon_I =|dI^{1/2}/dt|$ should be used for the Kolmogorov-like phenomenology in this case:\\
$$
E(k) \propto \varepsilon_I^{2/3} k^{-4/3}  \eqno{(27)}
$$
 %%%%%%%%%%%%%%% 10 %%%%%%%%%%%%%%%%%%
\begin{figure} \vspace{-2cm}\hspace{-1.3cm}\centering
\epsfig{width=.5\textwidth,file=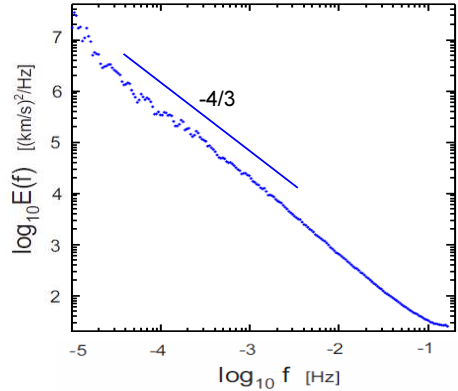} \vspace{-4.8cm}
\caption{Power spectrum of the bulk proton velocity field in the near-Earth orbit solar wind.} 
\end{figure}
%%%%%%%%%%%%%%%%%%%%%%%%%%%%%%%%%%% 
%%%%%%%%%%%%%%% 11 %%%%%%%%%%%%%%%%%%
\begin{figure} \vspace{-2cm}\hspace{-1.5cm}\centering
\epsfig{width=.5\textwidth,file=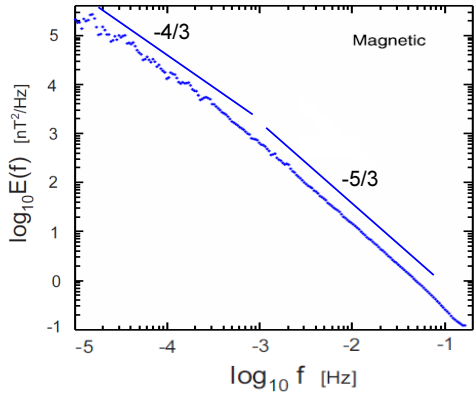} \vspace{-4.9cm}
\caption{As in the Fig. 10 but for the corresponding magnetic field.}
\end{figure}
%%%%%%%%%%%%%%%%%%%%%%%%%%%%%%%%%%%  
 %%%%%%%%%%%%%%% 12 %%%%%%%%%%%%%%%%%%
\begin{figure} \vspace{-0.5cm}\hspace{-1cm}\centering
\epsfig{width=.49\textwidth,file=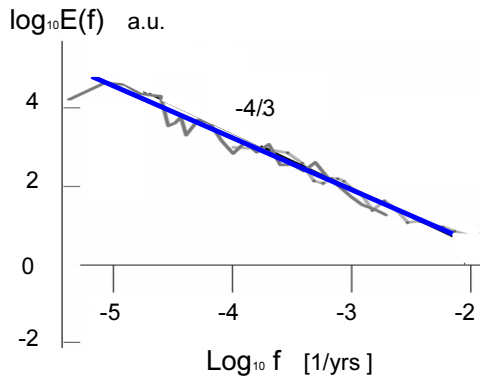} \vspace{-5cm}
\caption{Power spectrm of the paleo-temperature fluctuations.} 
\end{figure}
%%%%%%%%%%%%%%%%%%%%%%%%%%%%%%%%%%% 

   The Kolmogorov-like spectra Eq. (25) for magnetic field were observed at 1 AU for different phases of solar cycle (see, for instance, Refs. \cite{mg},\cite{bs} and references therein). However, as it follows from the previous consideration, simultaneous spectra for bulk velocity and for magnetic field can provide a more valuable information. In the Ref. \cite{prg} the power spectra of the proton bulk velocity and corresponding magnetic field in the solar wind were computed using the data obtained by in situ measurements at 1 AU (in the ecliptic plane ) acquired by the Wind spacecraft. The measurements were made near solar minimum: from 23 May to 16 July 1995. Figures  10 and 11 show the kinetic and magnetic energy spectra (the spectral data were taken from Fig. 6 of the Ref \cite{prg}). \\

     The straight lines in the Figs. 10 and 11 are drawn in order to indicate the spectral laws Eqs. (25) and (27) (in the log-log scales). One can see that for the kinetic energy spectrum the inertial range of scales is dominated by the helical scaling Eq. (27), while for the magnetic energy spectrum there are two subranges of the inertial range: the large-scales subrange is dominated by the helical scaling Eq. (27) and the small-scale subrange is dominated by the Kolmogorov-like scaling Eq. (25). \\
     
\section{Helical scaling in interstellar MHD turbulence} 

  Figure 12 shows the power spectrum of the paleo-temperature fluctuations obtained at Summit Greenland in the frames of the Greenland Ice Core Project - GRIP (the spectral data were taken from Fig. 1 of Ref. \cite{ls}). The black curve corresponds to the lower resolution GRIP core data which were interpolated to 200 year resolution and span the period back 240 kyr. The light gray curve corresponds to the mean of the GRIP 5.2 resolution data for the period of the last 90 kyr. \\
     
     For turbulent processes in the Heliosphere and on the Earth, the scaling observed in Fig. 12 (the straight line, see below) with such large time scales cannot exist. It is possible, however, if the scaling has a Galactic (interstellar) turbulence origin. The galactic cosmic rays (predominantly the high-energy protons) are the dominant source of penetrating ionizing particles in the atmosphere and have a significant influence on the amount of the global cloud cover, providing the condensation nuclei on which the cloud droplets form \cite{sf}-\cite{du}. Since low clouds have a large cooling effect on climate a relative abundance of the galactic cosmic rays results in more low clouds and, consequently, in lower temperatures. 

   On the other hand, the galactic cosmic ray intensity is modulated by the galactic magnetic field fluctuations. The galactic cosmic rays are gyrating along the field lines and are advected with the moving interstellar plasma. For the application of the above-mentioned Taylor hypothesis one can take the velocity of the solar system in relation to the cosmic microwave background rest frame, i.e. $|\langle {\bf V} \rangle| \simeq 370$ km/sec. In this case, the power spectrum of the galactic cosmic ray intensity in the vicinity of the Earth orbit can be approximately proportional to the galactic magnetic energy spectrum at this location and the scaling observed in the spectrum shown in Fig. 12 (the straight line corresponding to the Eq. (27)) can be used as an indication of the corresponding helical scaling in the spectrum of the interstellar magnetic field energy.

\end{document}